\documentclass[12pt]{article}
\usepackage{graphics}
\usepackage{graphicx}
\usepackage{epsfig}
\usepackage{wrapfig}

\begin{document}

\sloppy
\begin{flushright}{SIT-HEP/TM-20}
\end{flushright}
\vskip 1.5 truecm
\centerline{\large{\bf Brane Q-ball, branonium and brane inflation}}
\vskip .75 truecm
\centerline{\bf Tomohiro Matsuda
\footnote{matsuda@sit.ac.jp}}
\vskip .4 truecm
\centerline {\it Laboratory of Physics, Saitama Institute of
 Technology,}
\centerline {\it Fusaiji, Okabe-machi, Saitama 369-0293, 
Japan}
\vskip 1. truecm
\makeatletter
\@addtoreset{equation}{section}
\def\theequation{\thesection.\arabic{equation}}
\makeatother
\vskip 1. truecm

\begin{abstract}
\hspace*{\parindent}
We study the stability of branonium.
Contrary to the previous arguments, global structure of branonium is
not stable against spatial fluctuations.
We show that branonium decays into local objects, which looks like
 Q-balls in the effective action.
Therefore, we consider the idea of brane Q-balls in this paper.
Brane Q-balls are produced as the remnants of unstable 
branonium.
Although the brane Q-balls are conceptually different from the conventional
Q-balls, they look similar in some parametric regions.
We show how to distinguish brane Q-balls from conventional Q-balls.
\end{abstract}

\newpage
\section{Introduction}
The most promising scenario of unified theory will be included in string
theory, where the consistency is ensured by the requirement of
additional dimensions and supersymmetry.
Because of supersymmetry, many flat directions appear in the
effective Lagrangian.
Some flat directions parameterize the compactified
space of the string theory, or the 
distances between branes.
Brane distances play important roles in our arguments.\footnote{The idea
of large extra dimension may solve or weaken the hierarchy 
problem\cite{Extra_1}, and the most natural embeddings of this picture
in the string theory context would be realized by the brane
construction.
Inflation with a low 
fundamental scale ($M_* \ll M_p$) is discussed in ref.\cite{low_inflation,
matsuda_nontach}.
Baryogenesis with a low fundamental scale is discussed in
\cite{low_baryo} and \cite{defect-baryo-largeextra}.}
In this paper, we examine the stability of branonium\cite{Branonium}.
Contrary to the previous arguments,\footnote{
In ref.\cite{Ellison} it is discussed that the orbital motion of a light
anti D6-brane, taking account of possible time-variations in the
background moduli fields, the Coulomb-like central potential arising
through brane-antibrane interactions is modified to include
time-dependent prefactors, which generally preclude the existence of
stable elliptical orbits.} branonium is
not stable against spatial fluctuations and decays into local objects.
In this paper, we advocate a new idea of brane Q-balls.
The brane Q-balls are produced as the remnants of unstable branonium. 
Although the brane Q-balls are conceptually different from the conventional
Q-balls, they look similar in some parametric regions.
In this paper, we show why it is possible to distinguish the brane
Q-balls from the conventional Q-balls. 
When the charge of the brane Q-ball exceeds a critical value, its decay is
dominated by the radiation into the bulk.

\section{Stability of branonium and brane Q-balls}
In generic cases, branes have moduli, which allow relative motion in the
bulk. 
When considering the cosmology induced by such situation, non-static
configuration of these branes is therefore an interesting generic 
feature which deserves systematic study\cite{matsuda_nontach,
defect-baryo-largeextra, Brane-inflation, Dvali_alice, incidental_matsuda,
brane_defect_etc}.
``Branonium'' is essentially the dynamics of a light IIA or IIB anti
D$p$-brane in the background of a heavy D$p$-brane. 
For the case of D6-branes, it was shown\cite{Branonium}
that the effective world volume action in such background has rather
unusual properties.  
Their conclusion in ref.\cite{Branonium} was that the fully
relativistic, non-linear  
action allows for the existence of non-precessing, elliptical orbits for
the D6 anti-brane similar to those admitted for a non-relativistic
particle experiencing a central potential. 
The stability of these orbits against fluctuations and radiation
emission was also computed in ref.\cite{Branonium}.
The conclusion was that both effects could be minimized at large
distances. 
This planetary system of branes was named branonium.

On the other hand, in the usual arguments for the cosmological
formation of Q-balls, it is well known that tiny spatial
fluctuation effectively grows, which allows efficient formation of
the Q-balls from the almost homogeneous initial state\cite{Q-ball-simulation}. 

Obviously, the effective action of branonium looks similar to the one
that was used for the discussion about the efficient formation of the Q-balls. 
Then a naive question arises.
 ``Why branonium is stable? Doesn't it
fragment into local structure like Q-balls, as was discussed in the
conventional arguments?''
The purpose of this paper is to show why these two possibilities  
(branonium and
Q-ball) are contradicting each other. 
Our conclusion is that branonium is not stable against spatial fluctuations, but
decays into local structures that look like Q-balls in the effective action. 
The remnants of branonium look similar to the conventional Q-balls, 
which we call ``brane Q-balls''.
In spite of its name, the brane Q-ball is conceptually different from
the conventional Q-ball.

In the discussion about the stability of branonium\cite{Branonium},
the angular momentum $l$ of a brane in motion is assumed to be a
homogeneous parameter of the Universe. 
On the other hand, in the discussions about the formation of the
conventional Q-balls, the angular velocity $\omega$ is assumed to be
the homogeneous parameter. 
Obviously, these two different assumptions lead to distinctive results.
The latter assumption seems quite appropriate if one wants to discusses the
growth of the spatial fluctuation, as was discussed in ref.\cite{Q-ball,
Q-ball-simulation, Q-ball_futsuu_decay}.\footnote{Here the angular momentum
of branonium corresponds to the charge density of the conventional Q-ball.}
On the other hand, if one uses the former assumption, the fluctuation of
the brane distance induces the spatial gradient in $\omega$.\footnote{Imagine
a pair of oppositely charged strings rotating around each other.
Because of the conservation of the angular momentum, some parts will 
bulge but other parts will dwindle.
In this case, the angular momentum cannot be homogeneous because
the fluctuation of the angular velocity (or the fluctuation of the
distance) induces the movement of
the angular momentum.
As a result, the angular velocity becomes approximately the homogeneous
parameter of the system, while the angular momentum fragments, as was
discussed for conventional Q-balls.}
Then the former assumption leads to the structure of the
unnatural local windings that cost huge energy when the spatial
fluctuation grows. 
As a result, one might conclude that the branonium is stable against spatial
fluctuations, because it costs huge energy for the spatial fluctuations to grow.
In this case, however, it is \`a
priori impossible to examine the spatial fragmentation of $l$.
Thus in this case, the traditional arguments are trustworthy when one
wants to examine the stability of the branonium-like structure.

Here we consider another question.
In ref.\cite{Branonium}, the non-perturbative brane decay \`a la
Callan-Maldacena\cite{Callan-Maldacena} is discussed to be interpreted
as the sphaleron associated 
to the decay of the brane-antibrane system.
The conclusion in ref.\cite{Branonium} was that the nucleation
probability of the sphaleron is exponentially suppressed and negligible.
If such suppression is effective, nucleation rate of the true vacuum is
so small that branonium cannot decay into brane Q-balls.
Is this argument really true in realistic cases? 
Our answer is negative to the conclusion in ref.\cite{Branonium}. 
It is already well known that the conventional sphalerons in the
standard model are actually activated in the domain of the small (or
vanishing) Higgs expectation value, while
the sphaleron transition is suppressed in the vacuum of the large Higgs vev.
In the domain of the vanishing Higgs vev, where the sphalerons become
``massless'', the energy barrier does not make sense.
In our case, the Higgs field in the conventional sphaleron model is
corresponding to 
the distance between branes.
If the spatial fluctuation of the distance grows, the distance 
finally becomes negligible in a local domain.
Thus, one can easily understand why the Callan-Maldacena sphalerons are
activated.

According to the above arguments, it is natural to think that the brane
Q-ball, which is induced 
by a brane rotating at a distant around source branes, is a more generic
cosmological configuration than branonium. 
In the effective action, the angular momentum $l$ for the
configuration of branonium is corresponding to the charge of the Q-ball.  
The assumption of the homogeneous $l$ is not reliable, because the
effective Lagrangian is essentially the same as the ones that were
used for the discussion about the conventional
Q-balls\cite{Q-ball-simulation}, in which the  
rapid formation of the Q-balls is confirmed.
Our considerations are traditional in the arguments about
the Q-balls, where the angular velocity $\omega$ 
is assumed to be the homogeneous constant.\footnote{Here we define 
$\omega$ so that it is consistent with the definition of $l$ in
ref.\cite{Branonium}.}  

Although the effective action for the brane Q-ball looks similar to the
one for the conventional Q-balls, it is natural to expect that some
distinctive features will appear in the regions where the brane dynamics
becomes effective.  
In the followings, we examine if there are crucial discrepancies
between them. 

First, we consider the core structure of a large Q-ball.
In the core, the scalar field $\varphi$ that forms the Q-ball develops
large expectation value along a flat 
direction\cite{flat_Qball, Kasuya-Kawasaki}. 
In our case, as it can be seen from the effective Lagrangian, the brane
distance $\delta_{brane}$ is corresponding to the
scalar field $\varphi$ for the conventional Q-ball.
For the brane Q-ball, we define the scalar field $|\phi|\equiv
\delta_{brane} 
M_*^2$, where $M_*$ denotes the fundamental scale.
The rotation is defined by $\phi=|\phi| e^{i\omega t}$.\footnote{In this
paper, we 
assume that the direction of the rotational axis is homogeneous even if
$d$ (the number of the extra dimensions perpendicular to the rotating
brane) exceeds 2.
We think further discussions about the inhomogeneous rotational axis are
useful and interesting for the quantitative discussions about $d>2$ models. 
However, such analyses require numerical methods, and are obviously 
beyond the scope of this paper.
Although the quantitative discussions will be much more difficult than the 
conventional ones, the qualitative results will be consistent with the
$d=2$ cases.}  
It is easy to see that the potential becomes flat in the core, where
branes are placed at a distance.
In this region, our arguments are not different from the ones for the
conventional 
Q-balls.\footnote{Here we assume that the conventional Q-balls are
induced by the flat directions of the Minimal Supersymmetric 
Standard Model(MSSM)\cite{Q-ball}.}
Since the potential is nearly flat in the core, we can consider the
situation $\omega^2 \gg m^2_\phi$, where
$m_\phi$ denotes the effective mass of the scalar field $\phi$. 
Here we consider the most familiar example, the effective
action for the parallel N coincident source BPS D3-branes and a
rotating anti-D3-brane with $d$ large extra dimensions.\footnote{
Although we are considering D3-branes in this paragraph, it is important
to note about brane Q-balls constructed from higher-dimensional
branes. 
As our discussions about brane Q-balls are based on the effective action
of such brane configurations, the ``dimensionality'' should be found in
the effective action.
Therefore we can focus our attention to the dimensionality of the brane
configuration that appears in the effective action, because discussions
about Q-balls are based on the effective action.
Obviously, the potential of the brane distance $\phi$ depends on the
dimensionality of the branes.
However, the precise form of the potential is not relevant for the actual
discussion about Q-balls, since the only requirement that is used to
discuss the properties of ``fat'' Q-balls is the flatness of
the potential at large $\phi$.
The parameters of the ``fat'' Q-balls do not depend on the precise
form of the potential, as we will discuss in the later paragraph. 

On the other hand, we need more discussions about the kinetic term of
the scalar field $\phi$.
Here we define the coordinates of the compactified and uncompactified
space as $x^\mu$ and $x^a$, respectively.
Naively one might think that the kinetic term of the scalar field $\phi$
contains differentials of the form $f(\partial \phi/\partial x^a,..)$,
where $f$ is a function of $\partial \phi/\partial x^a$ and other
differentials and fields, so that
the discussion of the Q-balls must depend on the dimensionality of the
original branes. 
However, as was noted in ref.\cite{brane_defect_etc}, the effect of
compactification is significant for such cosmological fluctuations.
Since the compactification radius must be small compared to the
horizon size during inflation, any variation of a field in the
compactified direction is suppressed.
In our case, the length scale of the fluctuation of the scalar field
$\phi$ is much larger than the compactification radius.
Therefore the scalar field $\phi$ is constant in the compactified
directions, which makes our discussions about the dimensionality of
brane Q-balls much simpler than expected.
As a result, as far as the effective action is concerned, the
characteristic features of brane Q-balls do not 
depend on the dimensionality of the original brane
configuration(branonium), except for the bulk radiation that will
be discussed later in this paper.
Therefore it is important to note that we are not considering D3-branes as
a very special example, but as a minimal configuration
that represents the generic feature of such configurations.}
Since we are considering the brane Q-ball that might have been produced
after brane 
inflation, it is useful to consider the same situation as the model for
brane inflation\cite{Brane-inflation}.
In our simplest example, the effective action is described by;
\begin{equation}
\label{action_1}
S  \simeq -T_3 \int d^{4}\zeta 
\frac{1}{2}\partial^{\mu}X_a \partial_{\mu}X_a+...,
\end{equation}
where $X_a$ denotes the transverse coordinates of the brane, which represents
the relative distance if one assumes 
approximate spherical symmetry.
Then one can obtain the kinetic term for the field $\phi$,
\begin{eqnarray}
\label{action_2}
S &\simeq& -T_3 \int d^4 \zeta \frac{1}{2} 
\partial^{\mu}X_a \partial_{\mu}X_a\nonumber\\
&\equiv&-\int d^4 \zeta \frac{1}{2} 
\partial^\mu \phi \partial_\mu \phi.
\end{eqnarray}
If the two branes are separated at a distance, the potential for the
normalized field $|\phi|=\sqrt{T_3} \delta_{brane}$ is given
by\cite{Brane-inflation} 
\begin{equation}
\label{potential_1}
V(\phi) = M_*^4 \left[1-\frac{k M_*^4}{\phi^4}\right],
\end{equation}
where $M_*^4 \simeq T_{D3}$ is assumed, and $k$ is a constant of
$k<O(10^{-3})$.
Since the potential (\ref{potential_1}) is flat for $\delta_{brane} \gg
M_*^{-1}$, 
the situation in the core is quite similar to the usual Q-ball
configurations with the flat directions\cite{flat_Qball}.

To calculate the physical parameters for the brane Q-balls, we review the
discussion about the conventional fat 
Q-balls\cite{flat_Qball}, which are fat in a sense that the effective mass
$m_{\phi}$ in the core is much smaller than the one in the true
vacuum.
If the supersymmetry breaking in the true vacuum is induced by the gauge
mediation, flat directions of MSSM are naturally the candidates of
such fat Q-balls.
The fat Q-balls has two distinctive regions.
In the core of the Q-ball, where the potential is flat, the term
$\sim \omega^2 \phi^2$ dominates the potential. 
In this region the configuration becomes thick, which constitutes the
fat core.
Since the volume scales as the third power of the radius while the
surface area scales as the second power, its core rather than its
surface dominates the energy of the fat Q-ball.
The fat Q-ball has the following
properties\cite{Kasuya-Kawasaki} 
\begin{eqnarray}
\label{Q-ball-flat}
r_Q \simeq \frac{Q^{1/4}}{M_*}, && \omega \simeq \frac{M_*}{Q^{1/4}}\nonumber\\
\phi_Q \simeq M_* Q^{1/4},&& E_Q \simeq M_* Q^{3/4}
\end{eqnarray}
where $Q$, $r_Q$, and $E_Q$ denote the charge, the radius, and the
 energy of the Q-ball. 
Here $\phi_Q$ denotes the vacuum expectation value of $\phi$ in the core.
In the core, the brane Q-balls look quite similar to the conventional Q-balls.

Second, we examine the surface of the brane Q-balls.
Conventional fat Q-balls have the thin-wall regime, which 
interpolates between the fat core and the outer space of the true vacuum.
The potential is no more flat in this surface region, since the gauge
mediation of the supersymmetry breaking becomes effective.
In our model, similar to the conventional fat Q-balls, the brane
distance gradually becomes shorter toward the surface region.
At the same time, the effective potential for the field $\phi$
becomes steeper because of the tachyonic instability.
In the surface region where the effective mass $m_{\phi}$ exceeds
$|\omega|$, the 
approximation of the flat potential fails.
Because $\omega^2$ is much smaller than $m^2_\phi$ in this region,
one can ignore the time dependence and construct the steep
boundary configuration of the curved brane.
The boundary configuration looks similar to the conventional sphalerons.
Such a solution is already constructed in 
ref.\cite{Callan-Maldacena}.\footnote{See also ref.\cite{hashimoto}.} 
Here we do not repeat the arguments, since 
the exact solution is not needed in our case.
In ref. \cite{Callan-Maldacena, hashimoto}, it is shown
that the decaying brane solution
diverges at a critical radius.
If the configuration crosses the divergence in the surface of the brane
Q-ball, the surface effect will dominate. 
On the other hand, if the radius of the brane Q-ball is so large 
that the surface configuration does not induce such divergence, one can safely 
use the fat core approximation.
For the effective potential (\ref{potential_1}), the
surface region appears where the field $\phi$ becomes smaller than
$\phi_{surface}=k^{1/6} Q^{1/12}M_*$. 
In this region, the curved brane configuration is described by the
sphaleron-like configuration in
ref.\cite{Callan-Maldacena, hashimoto}.
In ref.\cite{Callan-Maldacena} and \cite{hashimoto}, the effective action for a
Dp-brane in the flat target space with a nontrivial background is
written, and the equation of motion is obtained for the transverse
coordinate $X$.\footnote{The equation of motion is given by eq.(3.6) in
ref.\cite{Callan-Maldacena}, or by eq.(2.11) in ref.\cite{hashimoto}.
Then the solution is given by eq.(3.9) and (3.10) in
ref.\cite{Callan-Maldacena}, or by eq.(2.13) in ref.\cite{hashimoto}.}
For us, the important characteristics of these solutions are the divergence
of $\partial_\mu X$ that appears at the critical radius
$r_c$.\footnote{The critical radius $r_c$ 
is given by the form $r_c \equiv(B^2-A^2)^{1/(2p-2)}$ in
ref.\cite{Callan-Maldacena}, and $r_c \equiv |p/2f|$ in
ref.\cite{hashimoto}.}
Although we will not review the explicit calculations in
ref.\cite{Callan-Maldacena}, it is easy to understand
that the divergences are inevitable in generic situations if the
nontrivial backgrounds are included. 
Thus, if the surface configuration of our brane Q-ball is crossing the
critical radius, the characteristics of the brane Q-ball will be determined
by the surface, not by the fat core.
According to ref.\cite{Callan-Maldacena, hashimoto},
one can see that the critical radius is physically at 
$r_c \simeq \delta_{brane}$.
In our model, $\delta_{brane}$ in the above condition is the distance
where the surface region starts. 
Thus, the critical radius is $r_c \simeq k^{1/6}Q^{1/12}
 M_*^{-1}$ in our model.\footnote{Of course, in the fat core of the
 brane Q-ball, the actual brane distance is 
different from $\delta_{brane}$ in the above condition.
$\delta_{brane}$ in the above condition is the brane distance where the
surface region starts. } 
As a result, the properties given in (\ref{Q-ball-flat}) are plausible only
when the brane Q-balls satisfy the condition $r_Q \gg r_c$.
Obviously, the above requirement is satisfied in our model.
There is no divergence in the surface region of the brane Q-ball.
The brane Q-balls look similar to the conventional fat Q-balls,
in a sense that their properties are given by eq.(\ref{Q-ball-flat}).

Third, we examine the decay channel of the brane Q-balls.
We show that the obvious distinction appears when the brane Q-ball
exceeds the critical charge.
In general, when one considers the time-dependent brane configurations, 
one cannot ignore the radiation into the bulk.
The radiation into the bulk is proportional to $a_b^2$, where $a_b$ is the
acceleration of the brane in motion in the bulk\cite{Branonium}.
We examine if this peculiar effect of the brane dynamics can exceed the
conventional decay mode of the Q-balls.
The power radiated into the bulk, $P= -\frac{dE_Q}{dt}$, is given
by\cite{Branonium} 
\begin{equation}
\label{radiation_into_bulk}
P \sim \frac{1}{8\pi} \left(\kappa_4 T_p V_p \right)^2 a_b^2,
\end{equation}
where $\kappa_4$ and $V_p$ are the 4-dimensional gravitational coupling
and the spatial volume of the Dp-brane, respectively.
Here we consider the local structure of the D3-brane.
The volume $V_p$ is given by the volume of the D3 brane Q-ball,
$V_3 \simeq 4\pi r_Q^3/3$.
We also assume the conventional form of the brane tension, $T_{Dp}
\simeq M_*^p$. 
On the other hand, the usual decay of the conventional Q-balls through
charged massless 
fermions has been studied in ref.\cite{Q-ball_futsuu_decay}, where an
upper bound is given by
\begin{equation}
\label{dtdq}
|\frac{dQ}{dt}| \leq \frac{\omega^3 A}{192 \pi^2}.
\end{equation}
Here $A=4\pi r_Q^2$ is the surface area of the Q-ball.
From eq.(\ref{dtdq}) and (\ref{Q-ball-flat}), we
obtain the upper bound
\begin{equation}
\label{conventional_decay}
\frac{dE_Q}{dt}\leq -c_q \frac{M_*^2}{Q^{1/2}},
\end{equation}
where the constant $c_q$ is $c_q \le O(10^{-3})$.
From eq.(\ref{Q-ball-flat}), (\ref{radiation_into_bulk}) and
(\ref{conventional_decay}), 
it is easy to see that the radiation into bulk dominates when
the charge $Q$ exceeds the critical value
\begin{equation}
\label{Q-ball_bulk_critical}
Q_c^{3/2} \simeq c_1\frac{M_p^2}{M_*^2},
\end{equation}
where $c_1\simeq 10^{-2}$.
Here the approximation $a_b \simeq \delta_{brane} \omega^2\simeq
(\phi_Q/M_*^2) \omega^2$ is used.  

The above results suggest that the brane Q-balls look similar to the
conventional Q-ball if it is in some parametric regions, however
a crucial difference appears when the charge exceeds the critical value.

Fourth, we consider the effect of the Hubble parameter.
In the followings, we show why the Hubble parameter is important for the
discussions about the stability of branonium.
As was discussed in ref.\cite{Q-ball-inflation}, the effect of the Hubble 
parameter is not negligible in the discussions about the conventional
Q-balls.
On the other hand, in the previous discussions about branonium, a
special limit is considered and the Hubble parameter is neglected. 
If the Hubble parameter is effective, the global structure of
branonium is nothing but the model for brane inflation with the
initial rotational motion\cite{Q-ball-inflation}. 
Let us consider a conventional example where the
time dependence 
of the energy density $\rho$ is given by the fluid equation
$\dot{\rho}=-3 \frac{\dot{a}}{a}(\rho+P)$,
where $P$ denotes the pressure of the Universe.
Adopting the general expressions for the energy density and pressure of
a homogeneous scalar field $\phi$, one can easily obtain the equation
\begin{equation}
\label{Equation_of_motion1}
\ddot{\phi}+3H\dot{\phi}=-\frac{dV}{d\phi}
\end{equation}
where $H$ is the Hubble parameter, which is given by the formula
\begin{equation}
\label{Hubble_eq}
H^2=\frac{1}{3M_p^2}\left[V(\phi) + \frac{1}{2}\dot{\phi}^2\right].
\end{equation}
If one wants to assume that the discussion in \cite{Branonium} is valid,
one should consider the following two constraints.
First, the field $\phi$ must violate the
slow-roll condition, which means $|V''(\phi)| \gg H^2$.
Otherwise, the configuration of branonium induces rapid expansion.
Here we have assumed that the Hubble parameter is $H^2\simeq
M_*^4/M_p^2$\cite{Brane-inflation}, as in the cases for the generic situations.
The above condition puts an upper bound for $\phi$.
In the case of D3-branes, the explicit form of the
condition becomes
\begin{equation}
\label{branonium_phi}
\phi \ll k^{1/6} M_* \times \left(\frac{M_p}{M_*}\right)^{1/3}.
\end{equation}
According to ref.\cite{Q-ball-inflation} and the above arguments, the
condition
$r_Q \gg H^{-1}$\cite{Q-ball-inflation} is required for the stable
configuration of branonium.
This condition is necessary to ensure the homogeneity of $l$ within the
Hubble radius. 
However, if the above constraint for the homogeneous $l$ is satisfied,
it is hard to believe that the condition (\ref{branonium_phi}) is
satisfied for the realistic values of $k <1$ and $M_* < M_p$. 

As a result, we conclude that the branonium configuration is a
precedent state of the brane Q-balls or brane inflation.
In this sense, branonium is not stable.
Branonium can appear only in a short transient period of the early Universe.

\section{Conclusions and Discussions}
In this paper, we have examined the stability of branonium.
Branonium is not stable, and decays into brane Q-balls.
The new idea of the brane Q-ball is studied.
The brane Q-ball looks similar to the conventional Q-ball when its charge 
is smaller than the critical value.
If it exceeds the critical value, the brane Q-ball decays
predominantly into the bulk, which is the most distinctive feature of
the brane 
Q-ball.

\section{Acknowledgment}
We wish to thank K.Shima for encouragement, and our colleagues in
Tokyo University for their kind hospitality.

\end{document}